\documentclass[epj]{svjour}
\usepackage{graphics}
\begin{document}
\title{A  symmetry for heavy nuclei: Proxy-SU(3)}

\author{\underline{Dennis Bonatsos}\inst{1} \and I. E. Assimakis \inst{1} \and N. Minkov\inst{2} \and Andriana Martinou\inst{1} \and R. B. Cakirli\inst{3} \and R. F. Casten\inst{4,5} \and
K. Blaum\inst{6}} % etc
%}                     % Do not remove
\institute{Institute of Nuclear and Particle Physics, National Centre for Scientific Research 
``Demokritos'', GR-15310 Aghia Paraskevi, Attiki, Greece
\and Institute of Nuclear Research and Nuclear Energy, Bulgarian Academy of Sciences, 72 Tzarigrad Road, 1784 Sofia, Bulgaria
\and Department of Physics, University of Istanbul, Istanbul, Turkey
\and Wright Laboratory, Yale University, New Haven, Connecticut 06520, USA
\and Facility for Rare Isotope Beams, 640 South Shaw Lane, Michigan State University, East Lansing, MI 48824 USA
\and  Max-Planck-Institut f\"{u}r Kernphysik, Saupfercheckweg 1, D-69117 Heidelberg, Germany}

\date{Received: date / Revised version: date}
\abstract{
The SU(3) symmetry realized by J. P. Elliott in the sd nuclear shell is destroyed in heavier shells 
by the strong spin-orbit interaction. However, the SU(3) symmetry has been used 
for the description of heavy nuclei in terms of bosons in the framework of the Interacting Boson 
Approximation, as well as in terms of fermions using the pseudo-SU(3) approximation. We introduce a new fermionic approximation, called the proxy-SU(3), and we comment on its similarities and differences with the other approaches.
 % end of PACS codes
} %end of abstract
\maketitle
\section{Introduction}
\label{intro}

The SU(3) symmetry has been introduced in nuclear structure by J. P. Elliott \cite{Elliott1,Elliott2},
who considered the sd shell nuclei and showed the microscopic origins of the connection 
between the nuclear quadrupole deformation and SU(3). A generalization of the Elliott SU(3) scheme to more than one nuclear shell has been obtained  in the framework of the microscopic symplectic model
\cite{Rosensteel}. Since then the SU(3) symmetry has been used in the framework of various 
algebraic models, especially for the study of medium-mass and heavy deformed nuclei, where the LS coupling scheme of the Elliott model breaks down \cite{Talmi}, while 
microscopic calculations are still out of reach. Descriptions in terms of bosons have been 
given in the frameowork of the Interacting Boson Model (IBM) \cite{IA} and of the Interacting Vector Boson Model (IVBM) \cite{Georgieva}, while fermionic descriptions have been provided by 
the Fermion Dynamical Symmetry Model (FDSM) \cite{FDSM}.
It underlies also the pseudo-SU(3) scheme \cite{pseudo1,pseudo2,DW1,DW2,Bahri68,Blokhin74,Ginocchio}, which we will discuss below, as well as the quasi-SU(3) symmetry \cite{Zuker1,Zuker2}, in which an approximate restoration of LS coupling in heavy nuclei is obtained, based on the smallness of certain $\Delta j=1$ matrix elements. 
  
On the other hand, many properties of heavy deformed nuclei have been successfully described in detail in terms of the Nilsson model \cite{Nilsson1,Nilsson2,RN}. Nilsson states 
 are labelled by $K[N n_z \Lambda]$, where $N$ is the number of oscillator quanta, 
$n_z$ is the number of quanta along the cylindrical symmetry axis, $\Lambda$ is the projection of the orbital angular momentum along the symmetry axis, and $K$ is the the projection of the total angular momentum along the symmetry axis, connected to $\Lambda$ by $K=\Lambda+\Sigma$, where $\Sigma$ is the the projection of the spin along the symmetry axis. For large deformations, the Nilsson wave functions reach the asymptotic limit, in which these quantum numbers become good quantum numbers, and they remain rather good even at intermediate deformation values \cite{Nilsson2}. 

Ben Mottelson  has remarked \cite{Mottelson} that  the asymptotic quantum numbers of the Nilsson 
model can be seen as a generalization of Elliott's SU(3), applicable to heavy deformed nuclei. Working in this direction, we have shown \cite{PRC1,PRC2} that a proxy-SU(3) symmetry of the Elliott type can be developed in heavy deformed nuclei. The development of the proxy-SU(3) scheme is based on
the so-called 0[110] pairs of Nilsson orbits related by $\Delta K [\Delta N \Delta n_z \Delta \Lambda]=0[110]$ \cite{Cakirli}. These pairs, which are characterized by high overlaps \cite{Karampagia}, have been shown to play a key role in the onset and development of nuclear deformation in the rare earth region \cite{Cakirli,Karampagia}. 

In the proxy-SU(3) scheme we also focus attention on  Nilsson 0[110] pairs, but in a different way. Instead of taking advantage of proton-neutron pairs, we use proton-proton and neutron-neutron pairs.  In this way we reveal an approximate SU(3) symmetry in heavy deformed nuclei, which can be used for predicting nuclear properties within the SU(3) symmetry using algebraic methods, as we shall see  in Refs. \cite{IoMartinou,IoSarant}. 

\section{The proxy-SU(3) model} 

%%%%%%%%%%%%%%%%%%%%%%%%%%% FIG. 1  %%%%%%%%%%%%%%%%%%%%%%%%%%%%%%%%%%%%%%%%%%

\begin{figure}[htb]
\resizebox{0.45\textwidth}{!}{%
  \includegraphics{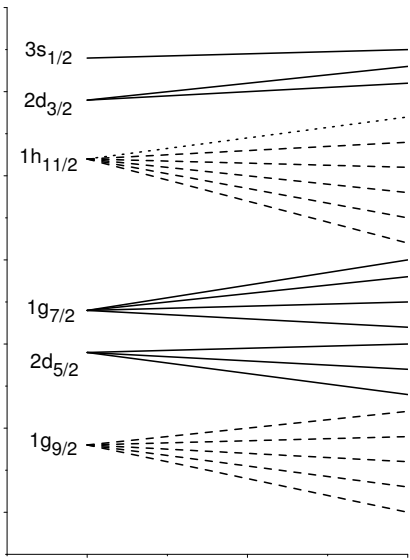}
}
\caption{Schematic representation of the 50--82 shell and the replacement leading to the proxy sdg shell.} 
 
\end{figure}

We are going to explain the basic idea behind the proxy-SU(3) scheme by considering as an example 
the 50-82 major nuclear shell, shown in Fig.~1. 

In the 50-82 major shell one finds the 3s$_{1/2}$, 2d$_{3/2}$, 2d$_{5/2}$, and 1g$_{7/2}$ orbitals (shown in Fig.~1 by solid lines). These are the pieces of the full sdg shell remaining after
the desertion  (because of the spin-orbit force) of the 1g$_{9/2}$ orbitals (indicated by dashed lines) into the next shell below, i.e. into the 28-50 nuclear shell. In addition, the 50-82 major shell contains the 1h$_{11/2}$ orbitals (shown by dashed lines plus one dotted line), 
which have invaded this shell from above, forced down and out of the pfh shell also by the spin-orbit force. 

The deserter 1g$_{9/2}$ orbital consists of the Nilsson orbitals 1/2[440], 3/2[431], 5/2[422], 7/2[413], 9/2[404].  These happen to be 0[110] partners of the 
1h$_{11/2}$ Nilsson orbitals 1/2[550], 3/2[541], 5/2[532], 7/2[523], 9/2[514], in the same order. 
Two orbitals being 0[110] partners possess exactly the same values of the projections of 
orbital angular momentum, spin, and total angular momentum, thus they are expected 
to exhibit identical behavior as far as properties related to angular momentum projections
are concerned. 0[110] partners have been first used in relation to proton-neutron pairs 
\cite{Cakirli}, found to correspond to increased strength of the proton-neutron interaction,
because of their large overlaps \cite{Karampagia}. 

One can thus think of replacing all of the  invading 1h$_{11/2}$ orbitals (the upper group of dashed lines in Fig.~1), except the 11/2[505] orbital (the dotted line in Fig.~1) in the 50-82 shell by their deserting 1g$_{9/2}$ counterparts (the lower group of dashed lines in Fig.~1), expecting nuclear properties related to angular momentum to be little affected, since angular momentum projections remain intact. However, one should take carefully into account that during this replacement the $N$ and $n_z$ quantum numbers have been changed by one unit each, thus
changing the sign of the parity. These changes will obviously affect the selection rules of various relevant matrix elements, as well as the avoided crossings \cite{Cejnar} in the Nilsson diagrams.
Detailed calculations to be shown in Ref.  \cite{IoAssim} will demonstrate that 
the changes inflicted in the Nilsson diagrams by these modifications are indeed minimal.
     
The 1h$_{11/2}$ 11/2[505] orbit has no 0[110] partner in the 1g$_{9/2}$ shell, thus it has been excluded from this replacement. However,  this orbit lies at the top of the 50-82 shell in the Nilsson diagrams \cite{Nilsson1,Nilsson2}, where it is  unlikely to find nuclei with large deformations.
The same remark applies to similar orbits in other shells such as the 13/2[606] orbit in the 82-126 shell.

After these two approximations have been performed, one is left with a collection of orbitals which form exactly the full sdg shell, which is known to possess a U(15) symmetry, having an SU(3) subalgebra \cite{BK}. However, in axially symmetric deformed nuclei the relevant symmetry is not spherical, but cylindrical \cite{Takahashi}. As  a consequence,  the relevant algebras are not U(N) Lie algebras, but more complicated versions of deformed algebras  \cite{RD,ND,PVI,Lenis,Sugawara,Arima}. Nevertheless, one can expect that some of the SU(3) features would appear within the approximate scheme. 

Since the present approximation scheme is based on the replacement of the invading from above abnornal parity orbitals (except the one with highest angular momentum) by their 0[110] counterparts deserting to the lower shell, with the latter being used as proxies of the former in subsequent considerations, we are going to call this approximation the {\sl proxy-SU(3)} model. 

The same approximation can be made  in the 28-50, 82-126, 126-184 shells,
 which thus become  approximate pf, pfh, sdgi shells, respectivel. These shells  are known to correspond  to U(10), U(21), U(28) algebras having SU(3) subalgebras (see \cite{BK} and references therein). 

\section{The pseudo-SU(3) scheme}

The present approach exhibits several similarities with and differences from the pseudo-SU(3) 
scheme, which has been extremely useful in the study of many properties of medium-mass  and heavy nuclei away from closed shells. We list here some of these studies.

1) Yrast \cite{DW1,DW2,Naqvi516} and non-Yrast \cite{Popa62,Popa69,Vargas70,Popa403,Vargas49,Vargas53} bands of even-even deformed nuclei.

2) Normal parity bands and excited bands in odd-mass nuclei \cite{Naqvi536,Vargas61,Vargas673,Vargas64,Vargas66}.

3) The scissors mode and magnetic dipole excitations \cite{Castanos180,Beuschel57,Rompf57,Draayer25,Beuschel61,Vargas551}.

4) Superdeformed bands \cite{Dudek59,Nazarewicz64}.

5) Double-beta decay \cite{Castanos571,Hirsch51,Hirsch589}, neutrinoless double-beta decay \cite{Hirsch582}, and double-electron capture \cite{Ceron471}. 

6) It should be pointed out that in the case of pseudo-SU(3) a unitary transformation connecting 
the normal parity orbitals to the pseudo-SU(3) space is known \cite{unitary1,unitary2}.

It is expected that using a large number of results regarding the study 
of fermionic systems  by algebraic techniques, already developed and used in the pseudo-SU(3) 
framework, one would be able to  perform further complementary studies using the proxy-SU(3) model.

\section{Conclusions}

A new approximate SU(3) symmetry applicable in heavy deformed nuclei has been suggested
\cite{PRC1,PRC2}, called the proxy-SU(3) scheme. In Ref.
\cite{IoAssim} a detailed numerical study will demonstrate the validity of the approximation, 
while in Refs. \cite{IoMartinou,IoSarant} some first applications of the method 
in predicting nuclear properties will be described.

\end{document}